\documentstyle[12pt]{article}
\begin{document}
\setlength{\unitlength}{0.7mm}\thicklines
\tolerance=2000
\hbadness=2000
\begin{flushright}
IHEP 94-51\\
21 April 1994
\end{flushright}
\begin{center}
$B_c$ spectroscopy\\
\vspace*{0.5cm}
{V.V.Kiselev, A.K.Likhoded, A.V.Tkabladze}\\
Institute for High Energy Physics,\\
Protvino, Moscow Region, 142284, Russia.
\end{center}

\begin{abstract}

In the framework of potential models for heavy quarkonium the mass spectrum
for the system ($\bar b c$) is considered. Spin-dependent splittings,
taking into account a change of a constant for effective Coulomb interaction
between the quarks, and widths of radiative transitions between the
($\bar b c$) levels are calculated. In the framework of QCD sum rules,
masses of the lightest vector $B_c^*$ and pseudoscalar $B_c$ states are
estimated, scaling relation for leptonic constants of heavy quarkonia is
derived, and the leptonic constant $f_{B_C}$ is evaluated.

\end{abstract}

\section*{Introduction}

Recently, theoretical interest has been rising to the study of
$B_c$ meson, the heavy ($\bar b c$) quarkonium with open charm and beauty.
This interest is stimulated by experimental $B_c$ search, being performed at
FNAL and LEP detectors.

{}From one side, similarly to $D$ and $B$ mesons with the open charm and
beauty, respectively, $B_c$ is long-living particle, decaying due to
the weak interaction. From another side, $B_c$ contains the heavy quarks, only,
and, therefore, it can be reliably described by the use of the methods,
developed for the ($\bar c c$) charmonium and the ($\bar b b$) bottomonium.

As for the $B_c$ production, exact analytic expressions have been recently
derived for functions of the heavy quark fragmentation into the heavy
quarkonium \cite{1b,2b,3b}, in the scaling limit $M^2/s \ll 1$.
The functions depend on the ratios of the $b$- and $c$-quark masses to
the $B_c$ mass. The normalization of the $b \to B_c^{(*)-} c$ fragmentation
functions also depends on the leptonic constant value for the $B_c$ meson.

In description of the $B_c$ decays, it is important to know its spectroscopic
characteristics.

Some preliminary estimates of the bound state masses of the ($\bar b c$)
system have been made in papers of refs.\cite{2k,5ka}, devoted to
the description of the charmonium and bottomonium properties, and in
ref.\cite{7ka}. Recently in refs.\cite{10k} and \cite{20k}, revised analysis
of the $B_c$ spectroscopy has been performed in the framework of the
potential approach and QCD sum rules.

In the present paper we consider the ($\bar b c$) spectroscopy with taking
into the account the change of effective Coulomb interaction constant,
defining spin-dependent splittings of the quarkonium levels. We calculate
the widths of radiative transitions between the levels and analyse the
leptonic constant $f_{B_C}$ in the framework of the QCD sum rules in the
scheme, allowing one to derive scaling relation for the leptonic constants
of the heavy quarkonia.

In Section 1 we calculate the mass spectrum of the ($\bar b c$) system
with the account of the spin-dependent forces. In Section 2 the widths
of the radiative transitions in the $B_c$ meson family are evaluated.
In Section 3 the leptonic constant of $B_c$ is calculated. In Conclusion
we discuss the obtained results.

\section{Mass Spectrum of $B_c$ Mesons. }

The $B_c$ meson is the heavy ($\bar b c$) quarkonium with
the open charm and beauty.
It occupies intermediate place in the mass spectrum of the heavy quarkonia
between the ($\bar c c$) charmonium and the ($\bar b b$) bottomonium.
The approaches, applied to study the charmonium and the bottomonium, can be
expanded to the description of the $B_c$ meson properties, and
experimental observation of $B_c$ could test these approaches and it could
be used for the detailed quantitative study of the mechanisms of the heavy
quark production, hadronization and decays.

In the present section we obtain the results on the $B_c$ meson spectroscopy.
We will show that below the threshold for the hadronic decay of the
($\bar b c$) system into the $BD$ meson pair, there are 16 narrow bound states,
decaying, by cascade way, into the lightest pseudoscalar $B_c^+(0^-)$  state
with the mass $m(0^-)\simeq 6.25$ GeV.

\subsection{Potential.}

The mass spectra of the charmonium and the bottomonium are studied
experimentally in details \cite{1k} and they are very well described
in the framework of the phenomenological potential models of the
nonrelativistic heavy quarks \cite{2k,5ka,3k,4k,5k}. To describe the mass
spectrum of the ($\bar b c$) system, one would prefer to use the potentials,
whose parameters do not depend on the flavours of the heavy quarks, composing
the heavy quarkonium, i.e. one would use the potentials, which rather
accurately
describe the mass spectra of ($\bar c c $) as well as ($\bar b b$), with one
and the same set of the potential parameters. The use of such potentials allows
one to avoid an interpolation of the potential parameters from the values,
fixed by the experimental data on the ($\bar c c$) and ($\bar b b$) systems,
to the values in the intermediate region of the ($\bar b c$) system.

As it has been shown in ref.\cite{6k}, with accuracy up to an additive shift,
the potentials, independent of the heavy quark flavours
\cite{2k,5ka,3k,4k,5k}, coincide each to other in the region of the average
distances between the heavy quarks in the ($\bar c c$) and ($\bar b b$)
systems, so
\begin{equation}
0.1\;fm < r < 1\;fm\;, \label{0}
\end{equation}
although those potentials have different asymptotic behaviour in the regions
of very low ($r\to 0$) and very large ($r\to \infty$) distances.

In Cornel model \cite{2k} in accordance with the asymptotic freedom in QCD,
the potential does behave Coulomb-likely at low distances, and the term,
confining the quarks, rises linearly at large distances
\begin{equation}
V_C(r) = -\frac{4}{3}\;\frac{\alpha_S}{r} + \frac{r}{a^2} + c_0\;,
\end{equation}
so that
\begin{eqnarray}
\alpha_S & = & 0.36\;, \nonumber \\
a & = & 2.34\;\; GeV^{-1}\;, \nonumber \\
m_c & = & 1.84\;\; GeV\;, \nonumber \\
c_0 & = & -0.25 \;\; GeV \;.
\end{eqnarray}

The Richardson potential \cite{3k} and its modifications in refs.\cite{5k} and
\cite{6ka} also respects to the behaviour, expected in the framework of QCD, so
\begin{equation}
V_R(r) = -\int\frac{d^3q}{(2\pi)^3}\;e^{i\vec{r}\vec{q}}\;
\frac{4}{3}\;\frac{12\pi}{27}\;\frac{1}{q^2 \ln(1+q^2/\Lambda^2)}\;,
\end{equation}
so that
\begin{equation}
\Lambda  =  0.398\;\; GeV\;.
\end{equation}

In the region of the average distances between the heavy quarks (\ref{0}),
the QCD-motivated potentials allow the approximations by the power
(Martin) or logarithmic potential.

The Martin potential has the form \cite{4k}
\begin{equation}
V_M(r) = -c_M + d_M (\Lambda_M r)^k\;, \label{m}
\end{equation}
so that
\begin{eqnarray}
\Lambda_M & = & 1\;\;GeV\;, \nonumber \\
k & = & 0.1\;, \label{2}\\
m_b & = & 5.174\;\; GeV\;, \nonumber \\
m_c & = & 1.8\;\; GeV\;, \nonumber \\
c_M & = & 8.064 \;\; GeV \;, \nonumber \\
d_M & = & 6.869 \;\; GeV\;.\nonumber
\end{eqnarray}

The logarithmic potential is equal to \cite{k11}
\begin{equation}
V_L(r) = c_L + d_L \;\ln (\Lambda_L r)\;, \label{l}
\end{equation}
so that
\begin{eqnarray}
\Lambda_L & = & 1\;\;GeV\;, \nonumber \\
m_b & = & 4.906\;\; GeV\;, \nonumber \\
m_c & = & 1.5\;\; GeV\;, \\
c_L & = & -0.6635 \;\; GeV\;, \nonumber \\
d_L & = & 0.733 \;\; GeV\;.\nonumber
\end{eqnarray}

The approximations of the nonrelativistic potential of the heavy quarks
in the distance region (\ref{0}) in the form of the power (\ref{m}) and
logarithmic (\ref{l}) laws, allow one to study its scaling properties.

In accordance with the virial theorem, the average kinetic energy of the quarks
in the bound state is determined by the following expression
\begin{equation}
<T> = \frac{1}{2}\; <\frac{r dV}{dr}>\;. \label{v}
\end{equation}
Then, the logarithmic potential allows one to conclude, that for the
quarkonium states one gets
\begin{equation}
<T_L> = const.\;, \label{t}
\end{equation}
independently of the flavours of the heavy quarks, composing the heavy
quarkonium,
$$const. = d_L/2 \approx 0.367\;\; GeV\;.$$

In the Martin potential, the virial theorem (\ref{v}) allows one to obtain
the expression
\begin{equation}
<T_M> = \frac{k}{2+k}\;(c_M+E)\;,
\end{equation}
where $E$ is the binding energy of the quarks in the heavy quarkonium.
Phenomenologically, one has $|E| \ll c_M$ (for example,
$E(1S, c \bar c) \simeq -0.5$ GeV),
so that, neglecting the binding energy of the heavy quarks inside the heavy
quarkonium, one can conclude that the average kinetic energy of the heavy
quarks is a constant value, independent of the quark flavours and the number
of the radial or orbital excitation.
The accuracy of such approximation for $<T>$ is about 10\%, i.e.
$|\Delta T/T| \sim 30\div 40$ MeV.

\begin{table}[b]
\caption{The Mass Difference for the Two Lightest Vector States of the
Different Heavy Systems,  $\Delta M = M(2S) - M(1S)$ in MeV.}
\label{t1}
\begin{center}
\begin{tabular}{||c|c|c|c|c||}
\hline
system & $\Upsilon$ & $\psi$ & $B_c$ & $\phi$ \\
\hline
$\Delta M$ & 563 & 588 & 585 & 660\\
\hline
\end{tabular}
\end{center}
\end{table}

{}From the Feynman-Hellmann theorem for the system with the reduced mass $\mu$,
one has
\begin{equation}
\frac{dE}{d\mu} = -\;\frac{<T>}{\mu}\;,
\end{equation}
and, in accordance with condition (\ref{t}), it follows that the difference
of the energies for the radial excitations of the heavy quarkonium levels
does not depend on the reduced mass of the $Q\bar Q'$ system
\begin{equation}
E(\bar n,\mu) - E(n,\mu) = E(\bar n,\mu') - E(n,\mu')\;.\label{e1}
\end{equation}

Thus, in the approximation of both the low value for the binding energy
of the quarks and zero value for the spin-dependent splittings of the levels,
the heavy quarkonium state density does not depend on the heavy quark flavours
\begin{equation}
\frac{dn}{dM_n} = const.\label{e2}
\end{equation}
The given statement has been also derived in ref.\cite{7k} by the use
of the Bohr-Sommerfeld quantization of the S-wave states for the heavy
quarkonium system with the Martin potential \cite{4k}.

Relations (\ref{e1})-(\ref{e2}) are phenomenologically confirmed for the
vector S-levels of the $b\bar b$, $c\bar c$, $s\bar s$ systems \cite{1k}
(see table \ref{t1}).

Thus, the structure of the nonsplitted S-levels of the ($\bar b c$) system
must not only qualitatively, but quantitatively repeat the structure
of the S-levels for the $\bar b b$ and $\bar c c$ systems, with the accuracy
by overall additive shift of the masses.

Moreover, in the framework of the QCD sum rules, the universality of the
heavy quark nonrelativistic potential (the independence on the
flavours and the scaling properties (\ref{t}), (\ref{e1}),
(\ref{e2})) allows one to obtain the scaling relation for the
leptonic constants of the S-wave quarkonia \cite{7k}
\begin{equation}
\frac{f^2}{M} = const.\;,\label{law1}
\end{equation}
independently of the heavy quark flavours in the regime, when
$$  |m_Q-m_{Q'}|\;\; is\;\;restricted\;, \;\;\Lambda_{QCD}/m_{Q,Q'} \ll 1\;,$$
i.e., when one can neglect the heavy quark mass difference, and, in the regime,
when the mass difference is not low, one has
\begin{equation}
\frac{f^2}{M}\;\biggl(\frac{M}{4\mu}\biggr)^2 = const.\;, \label{law}
\end{equation}
where
$$\mu = \frac{m_Q m_{Q'}}{m_Q+m_{Q'}}\;.$$

Consider the mass spectrum of the ($\bar b c$) system with the Martin
potential \cite{4k}.

Solving the Schr\" odinger equation with the potential (\ref{m})  and
the parameters (\ref{2}), one finds the $B_c$ mass spectrum and the
characteristics of the radial wave functions $R(0)$ and $R'(0)$,
shown in the tables \ref{t2} and \ref{t3}, respectively.

\begin{table}[t]
\caption{The Energy Levels of the $\bar b c$  System, Calculated with no
Taking into the Account Relativistic Corrections, in GeV.}
\label{t2}
\begin{tabular}{|c|c|c|c||c|c|c|c||c|c|c|c|}
\hline
n & \cite{7ka} & \cite{8k} & \cite{6ka} &
n & \cite{7ka} & \cite{8k} & \cite{6ka} &
n & \cite{7ka} & \cite{8k} & \cite{6ka} \\
\hline
1S & 6.301 & 6.315 & 6.344 & 2P & 6.728 &6.735&6.763 & 3D & 7.008
&7.145  &7.030  \\
2S & 6.893 & 7.009 & 6.910 & 3P & 7.122 & -- &7.160 & 4D & 7.308
&  --& 7.365 \\
3S & 7.237 & --    & 7.024 & 4P & 7.395 & -- & -- & 5D & 7.532 &  --&  --\\
\hline
\end{tabular}
\end{table}

\begin{table}[b]
\caption{The Characteristics of the Radial Wave Functions $R_{nS}(0)$
(in GeV$^{3/2}$) and $R'_{nP}(0)$ (in GeV$^{5/2}$),
Obtained from the Schr\" odinger Equation.}
\label{t3}
\begin{center}
\begin{tabular}{||c|c|c||}
\hline
n & Martin  & ~~~~\cite{10k}~~~~\\
\hline
$R_{1S}(0)$  & 1.31 & 1.28 \\
$R_{2S}(0)$  & 0.97 & 0.99 \\
$R'_{2P}(0)$ & 0.55 & 0.45 \\
$R'_{3P}(0)$ & 0.57 & 0.51 \\
\hline
\end{tabular}
\end{center}
\end{table}

The average kinetic energy of the levels, lying below the threshold for
the ($\bar b c$) system decay into the $BD$ pair, is presented in
table \ref{t4}, from which one can see that the term, added to the
radial potential due to the orbital rotation,
\begin{equation}
\Delta V_l =\frac{\vec{L}^2}{2\mu r^2}
\end{equation}
weakly influences on the value of the average kinetic energy, and the binding
energy for the levels with $L\neq 0$ is essentially determined by the orbital
rotation energy, that does not still strongly depend on the quark flavours
(see table \ref{t5}), so that the structure of the nonsplitted levels of the
($\bar b c$) system with $L\neq 0$ must quantitatively repeat the structure
of the charmonium and bottomonium levels, too.

\begin{table}[t]
\caption{The Average Kinetic and Orbital Energies of the Quark Motion in the
($\bar b c$) System, in GeV.}
\label{t4}
\begin{center}
\begin{tabular}{||c|l|l|l|l|l||}
\hline
nL          & 1S   & 2S   & 2P   & 3P   & 3D\\
\hline
$<T>$       & 0.35 & 0.38 & 0.37 & 0.39 & 0.39\\
$\Delta V_l$& 0.00 & 0.00 & 0.22 & 0.14 & 0.29\\
\hline
\end{tabular}
\end{center}
\end{table}

\begin{table}[b]
\caption{The Average Energy of the Orbital Motion in the Heavy Quarkonia,
in the Model with the Martin Potential, in GeV.}
\label{t5}
\begin{center}
\begin{tabular}{||c|l|l|l||}
\hline
system        & $\bar c c$ & $\bar b c$& $\bar b b$\\
\hline
$\Delta V_l(2P)$& 0.23       & 0.22      & 0.21\\
\hline
\end{tabular}
\end{center}
\end{table}

\subsection{Spin-dependent Splitting of the ($\bar b c$) Quarkonium.}

In accordance with the results of refs.\cite{8k,9k},
to take into the account the spin-orbital and spin-spin interactions,
causing the splitting of the $nL$-levels ($n$ is the principal quantum number,
$L$ is the orbital momentum), one introduces the additional term to the
potential, so it has the form
\begin{eqnarray}
   V_{SD}(\vec{r}) & = &\biggl(\frac{\vec{L}\cdot\vec{S}_c}{2m_c^2} +
\frac{\vec{L}\cdot\vec{S}_b}{2m_b^2}\biggr)\;
\biggl(-\frac{dV(r)}{rdr}+\frac{8}{3}\;\alpha_S\;\frac{1}{r^3}\biggr)
+ \nonumber \\
{}~ & ~ & +\frac{4}{3}\;\alpha_S\;\frac{1}{m_c m_b}\;
\frac{\vec{L}\cdot\vec{S}}{r^3}
+\frac{4}{3}\;\alpha_S\;\frac{2}{3m_c m_b}\;
\vec{S}_c\cdot\vec{S}_b\;4\pi\;\delta(\vec{r}) \label{3} \\
{}~ & ~ & +\frac{4}{3}\;\alpha_S\;\frac{1}{m_c
m_b}\;(3(\vec{S}_c\cdot\vec{n})\;
(\vec{S}_b\cdot\vec{n}) - \vec{S}_c\cdot\vec{S}_b)\;\frac{1}{r^3}\;,
\;\;\vec{n}=\frac{\vec{r}}{r}\;,\nonumber
\end{eqnarray}
where $V(r)$ is the phenomenological potential, confining the quarks,
the first term takes into the account the relativistic corrections to the
potential $V(r)$; the second, third and fourth terms are the relativistic
corrections, coming from the account of the one gluon exchange between the
$b$ and $c$ quarks; $\alpha_S$ is the effective constant of the quark-gluon
interaction inside the ($\bar b c$) system.

The value of the $\alpha_S$ parameter can be determined by the following way.

The splitting of the S-wave heavy quarkonium ($Q_1\bar Q_2$) is determined
by the expression
\begin{equation}
\Delta M(nS) = \alpha_S\;\frac{8}{9m_1m_2}\;|R_{nS}(0)|^2\;,
\end{equation}
where $R_{nS}(0)$ is the value of the radial wave function of the quarkonium,
at the origin. Using the experimental value of the S-state splitting in the
$c\bar c$ system \cite{1k}
\begin{equation}
\Delta M(1S,\;c\bar c) =117\pm2\;\;MeV\;,
\end{equation}
and the $R_{1S}(0)$ value, calculated in the potential model for the
$c\bar c$ system, one gets the model-dependent value of the
$\alpha_S(\psi)$ constant for the effective Coulomb interaction of the heavy
quarks (in the Martin potential, one has $\alpha_S(\psi) = 0.44$).

In ref.\cite{10k} the effective constant value, fixed by the described way,
has been applied to the description of not only the $c\bar c$ system, but
also the $\bar b c$ and $\bar b b$ quarkonia.

In the present paper we take into the account the variation of the effective
Coulomb interaction constant versus the reduced mass of the system ($\mu$).

In the one-loop approximation at the momentum scale $p^2$, the "running"
coupling constant in QCD is determined by the expression
\begin{equation}
\alpha_S(p^2) = \frac{4\pi}{b\ln(p^2/\Lambda_{QCD}^2)}\;,
\end{equation}
where $b=11-2n_f/3$, and $n_f=3$, when one takes into the account the
contribution by the virtual light quarks, $p^2 < m^2_{c,b}$.

In the model with the Martin potential, for the kinetic energy of the quarks
($c\bar c$) inside $\psi$, one has
\begin{equation}
<T_{1S}(c\bar c)> \simeq 0.357\;\;GeV\;,
\end{equation}
so that, using the expression for the kinetic energy,
\begin{equation}
<T> =\frac{<p^2>}{2\mu}\;,
\end{equation}
one gets
\begin{equation}
\alpha_S(p^2) = \frac{4\pi}{b\ln(2<T>\mu/\Lambda_{QCD}^2)}\;,\label{al1}
\end{equation}
so that $\alpha_S(\psi) = 0.44$ at
\begin{equation}
\Lambda_{QCD} \simeq 164\;\;MeV \;.\label{al2}
\end{equation}

As it has been noted in the previous section, the value of the kinetic energy
of the quark motion weakly depends on the heavy quark flavours, and it,
practically, is constant, and, hence, the change of the effective
$\alpha_S$ coupling is basically determined by the variation of the reduced
mass of the heavy quarkonium. In accordance with eqs.(\ref{al1})-(\ref{al2})
and table \ref{t4}, for the $(\bar b c)$ system one has
\begin{center}
\begin{tabular}{cccccc}
nL       & 1S    & 2S    & 2P    & 3P    & 3D\\
$\alpha_S$ & 0.394 & 0.385 & 0.387 & 0.382 & 0.383
\end{tabular}
\end{center}

Note, the Martin potential leads to the $R_{1S}(0)$ values, which, with the
accuracy up to $15\div20\%$, agrees with  the experimental values of the
leptonic decay constants for the heavy $c\bar c$ and $b\bar b$ quarkonia.
The leptonic constants are determined by the expression
\begin{equation}
\Gamma(Q\bar Q \to l^+l^-) = \frac{4\pi}{3}\;e_Q^2\;\alpha^2_{em}\;
\frac{f_{Q\bar Q}^2}{M_{Q\bar Q}}\;,
\end{equation}
where $e_Q$ is the heavy quark charge.

In the nonrelativistic model one has
\begin{equation}
f_{Q\bar Q} = \sqrt{\frac{3}{\pi M_{Q\bar Q}}}\;R_{1S}(0)\;.\label{27s}
\end{equation}
For the effective Coulomb interaction of the heavy quarks in the basic
1S-state one has
\begin{equation}
R_{1S}^{Coul}(0) = 2 \;\biggl(\frac{4}{3}\;\mu\;\alpha_S\biggr)^{3/2}\;.
\end{equation}
One can see from table \ref{t7}, that, taking into the account the variation
of the effective $\alpha_S$ constant versus the reduced mass of the heavy
quarkonium (see eq.(\ref{al1})), the Coulomb wave functions give the values
of the leptonic constants for the heavy 1S-quarkonia, so that in the framework
of the accuracy of the potential models, those values agree with the
experimental values and the values, obtained by the solution of the
Schr\" odinger equation with the given potential.

\begin{table}[t]
\caption{The Leptonic Decay Constants of the Heavy Quarkonia, the Values,
Measured Experimentally and Obtained in the Model with the Martin Potential,
in the Model with the Effective Coulomb Interaction and from the Scaling
Relation (*), in GeV.}
\label{t7}
\begin{center}
\begin{tabular}{||c|c|c|c|c||}
\hline
model    & exp.\cite{1k} & Martin    & Coulomb      &  * \\
\hline
$f_\psi$  & $410\pm15$     & $547\pm80$ & $426\pm60$ & $410\pm40$\\
$f_{B_C}$ & --             & $510\pm80$ & $456\pm70$ & $460\pm60$\\
$f_\Upsilon$ & $715\pm15$  & $660\pm90$ & $772\pm120$& $715\pm70$\\
\hline
\end{tabular}
\end{center}
\end{table}

The taking into the account the variation of the effective Coulomb interaction
constant becomes especially notable for the $\Upsilon$-particles, for which
$\alpha_S(\Upsilon)\simeq 0.33$ instead the fixed value $\alpha_S = 0.44$.

Thus, calculating the splitting of the ($\bar b c$) levels, we take into
the account the $\alpha_S$ dependence on the reduced mass of the heavy
quarkonium.

As one can see from eq.(\ref{3}), in contrast to the $LS$-coupling
in the $(\bar c c)$ and $(\bar b b)$ systems, there is the $jj$-coupling
in the heavy quarkonium, where the heavy quarks have different masses
(here, $ \vec{L}\vec{S}_c$ is diagonalized at the given $\vec{J}_c$
momentum, ($\vec{J}_c =\vec{L}+\vec{S}_c$, $\vec{J}=\vec{J}_c +\vec{S}_b$),
$\vec{J}$ is the total spin of the system).
We use the following spectroscopic notations for the splitted levels of the
($\bar b c$) system, -- $n^{2j_c}L_J$.

One can easily show, that independently of the total spin $J$ projection
one has
\begin{eqnarray}
|^{2L+1}L_{L+1}> & = & |J=L+1,\;S=1>\;, \nonumber \\
|^{2L-1}L_{L-1}> & = & |J=L-1,\;S=1>\;,\label{3.c}  \\
|^{2L+1}L_{L}> & = & \sqrt{\frac{L}{2L+1}}|J=L,\;S=1> +
                    \sqrt{\frac{L+1}{2L+1}}|J=L,\;S=0>\;,\nonumber \\
|^{2L-1}L_{L}> & = & \sqrt{\frac{L+1}{2L+1}}|J=L,\;S=1> -
                    \sqrt{\frac{L}{2L+1}}|J=L,\;S=0>\;,\nonumber
\end{eqnarray}
where $|J,\;S>$ are the state vectors with the given values of the total
quark spin $\vec{S} = \vec{S}_c+\vec{S}_b$, so that the potential terms of
the order of $1/m_c m_b$, $1/m_b^2$ lead, generally speaking, to the mixing
of the levels with the different $J_c$ values at the given $J$ values.
The tensor forces (the last term in eq.(\ref{3})) are equal to zero at
$L=0$  or  $S=0$.

One can easily show, that
\begin{equation}
3 (n^p n^q -\frac{1}{3}\delta^{pq}) S_c^p S_b^q =
\frac{3}{2} (n^p n^q -\frac{1}{3}\delta^{pq}) S^p S^q\;,
\end{equation}
since for the quark spin one has
\begin{equation}
S_Q^p S_Q^q + S_Q^q S_Q^p = \frac{1}{2} \delta^{pq}\;.
\end{equation}
The averaging over the angle variables can be represented in the form
\begin{equation}
<L,\;m| n^p n^q |L,\;m'> = a(L^p L^q + L^q L^p)_{mm'}+b \delta^{pq}\;,
\label{3l}
\end{equation}
where $\vec{L}$ are the orbital momentum matrices in the respective
irreducible representation.

Let us use the following conditions.

1) The normalization of the unit vector,
\begin{equation}
<n^p n^q> \delta^{pq} = 1\;.
\end{equation}

2) The orthogonality of the radius-vector to the orbital momentum,
\begin{equation}
n^p L^p = 0\;,
\end{equation}

3) The commutation relations for the angle momentum,
\begin{equation}
[L^p; L^q] = i\epsilon^{pql} L_l\;.
\end{equation}
Then one can easily find, that in eq.(\ref{3l}) one gets
\begin{eqnarray}
a & = & - \frac{1}{4\vec{L}^2-3}\;,\\
b & = & \frac{2\vec{L}^2-1}{4\vec{L}^2-3}\;.
\end{eqnarray}
Thus, (see also ref.\cite{11k})
\begin{equation}
<6 (n^p n^q -\frac{1}{3}\delta^{pq}) S_c^p S_b^q> =
- \frac{1}{4\vec{L}^2-3} (6(\vec{L} \vec{S})^2 + 3 (\vec{L} \vec{S}) -2
\vec{L}^2 \vec{S}^2)\;. \label{3.t}
\end{equation}

\begin{figure}[t]
\begin{center}
\begin{picture}(175,150)
\put(15,30){\line(1,0){20}}
\put(15,33){$1S$}
\put(36,25){\line(1,0){20}}
\put(36,32){\line(1,0){20}}

\put(57,21){$^{0^-}$}
\put(57,29){$^{1^-}$}

\put(15,89){\line(1,0){20}}
\put(15,92){$2S$}
\put(36,87){\line(1,0){20}}
\put(36,90){\line(1,0){20}}

\put(57,83){$^{0^-}$}
\put(57,87){$^{1^-}$}

\put(15,124){\line(1,0){20}}
\put(15,127){$3S$}

\put(70,73){\line(1,0){20}}
\put(70,76){$2P$}
\put(91,68){\line(1,0){20}}
\put(91,72){\line(1,0){20}}
\put(91,74){\line(1,0){20}}
\put(91,76){\line(1,0){20}}

\put(112,63){$^{0^+}$}
\put(112,75){$^{2^+}$}
\put(112,67){$^{1^+}$}
\put(112,71){$^{1'^+}$}

\put(70,112){\line(1,0){20}}
\put(70,115){$3P$}
\put(91,109){\line(1,0){20}}
\put(91,111){\line(1,0){20}}
\put(91,112){\line(1,0){20}}
\put(91,114){\line(1,0){20}}

\put(112,102){$^{0^+}$}
\put(112,114){$^{2^+}$}
\put(112,106){$^{1^+}$}
\put(112,110){$^{1'^+}$}

\put(70,140){\line(1,0){20}}
\put(70,143){$4P$}

\put(125,101){\line(1,0){20}}
\put(125,104){$3D$}
\put(146,099){\line(1,0){20}}
\put(146,100){\line(1,0){20}}
\put(146,101){\line(1,0){20}}
\put(146,102){\line(1,0){20}}

\put(167,099){$^{1^-}$}
\put(167,095){$^{3^-}$}
\put(167,091){$^{2^-}$}
\put(167,103){$^{2'^-}$}

\put(125,131){\line(1,0){20}}
\put(125,134){$4D$}

\put(10,0){\framebox(165,150)}
\put(0,0){$6.0$}
\put(10,50){\line(1,0){3}}
\put(0,50){$6.5$}
\put(10,100){\line(1,0){3}}
\put(0,100){$7.0$}
\put(0,150){$7.5$}
\put(10,115){\line(1,0){55}}
\put(10,115.3){\line(1,0){55}}
\put(120,115){\line(1,0){55}}
\put(120,115.3){\line(1,0){55}}
\put(130,117){$BD$ $threshold$}
\end{picture}
\end{center}
\caption{The Mass Spectrum of the $(\bar b c)$ System with the Account
of the Splittings.}
\label{f1k}
\end{figure}
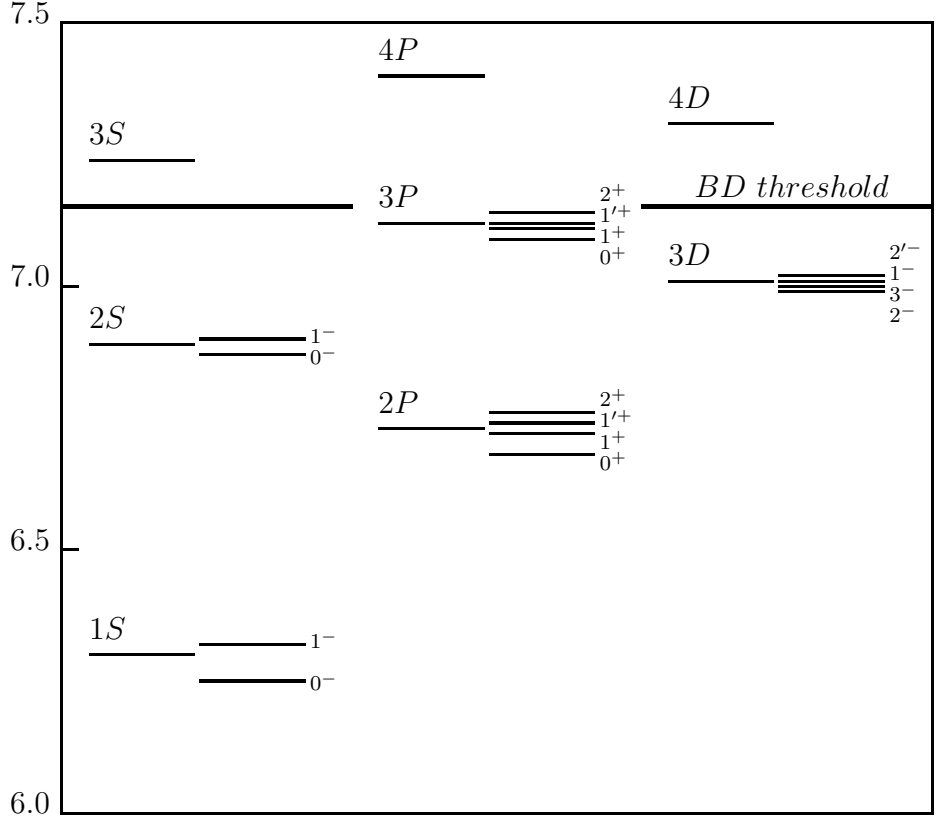

Using eqs.(\ref{3.c}), (\ref{3.t}), for the level shift, calculated in the
perturbation theory at $S=1$, one gets the following formulae
\begin{eqnarray}
   \Delta E_{n^1S_0} & = &- \alpha_S\;\frac{2}{3m_cm_b}\;|R_{nS}(0)|^2\;,\\
   \Delta E_{n^1S_1} & = & \alpha_S\;\frac{2}{9m_cm_b}\;|R_{nS}(0)|^2\;,
            \label{5}\\
   \Delta E_{n^3P_2} & = & \alpha_S\;\frac{6}{5m_cm_b}\;<\frac{1}{r^3}> +
\nonumber \\&&\frac{1}{4}\;\biggl(\frac{1}{m_c^2} +
\frac{1}{m_b^2}\biggr)\;
<-\frac{dV(r)}{rdr}+\frac{8}{3}\;\alpha_S\frac{1}{r^3}>\;, \\
   \Delta E_{n^1P_0} & = & - \alpha_S\;\frac{4}{m_cm_b}\;<\frac{1}{r^3}> -
\nonumber \\&&\frac{1}{2}\;\biggl(\frac{1}{m_c^2} +
\frac{1}{m_b^2}\biggr)\;
<-\frac{dV(r)}{rdr}+\frac{8}{3}\;\alpha_S\frac{1}{r^3}>\;,\\
   \Delta E_{n^5D_3} & = &  \alpha_S\;\frac{52}{21m_cm_b}\;<\frac{1}{r^3}> +
\nonumber \\&&\frac{1}{2}\;\biggl(\frac{1}{m_c^2} + \frac{1}{m_b^2}\biggr)\;
<-\frac{dV(r)}{rdr}+\frac{8}{3}\;\alpha_S\frac{1}{r^3}>\;,\\
   \Delta E_{n^3D_1} & = & - \alpha_S\;\frac{92}{21m_cm_b}\;<\frac{1}{r^3}> -
\nonumber \\&&\frac{3}{4}\;\biggl(\frac{1}{m_c^2} + \frac{1}{m_b^2}\biggr)\;
<-\frac{dV(r)}{rdr}+\frac{8}{3}\;\alpha_S\frac{1}{r^3}>\;,
\end{eqnarray}
where $R_{nS}(0)$ are the radial wave functions at $L=0$,  $<...>$  denote
the average values, calculated under the wave functions $R_{nL}({r})$.
The mixing matrix elements have the forms
\begin{eqnarray}
<~^3P_1|\Delta E|~^3P_1> & = & -\alpha_S\;\frac{2}{9m_cm_b}\;<\frac{1}{r^3}> +
\nonumber \\
 & &\biggl(\frac{1}{4m_c^2} - \frac{5}{12m_b^2}\biggr)\;
<-\frac{dV(r)}{rdr}+\frac{8}{3}\;\alpha_S\frac{1}{r^3}>\;,\\
<~^1P_1|\Delta E|~^1P_1> & = & -\alpha_S\;\frac{4}{9m_cm_b}\;<\frac{1}{r^3}> +
\nonumber \\
&&\biggl(-\frac{1}{2m_c^2} + \frac{1}{6m_b^2}\biggr)\;
<-\frac{dV(r)}{rdr}+\frac{8}{3}\;\alpha_S\frac{1}{r^3}>\;, \\
<~^3P_1|\Delta E|~^1P_1> & = &
-\alpha_S\;\frac{2\sqrt{2}}{9m_cm_b}\;<\frac{1}{r^3}>
\nonumber \\
&&-\frac{\sqrt{2}}{6m_b^2}\;
<-\frac{dV(r)}{rdr}+\frac{8}{3}\;\alpha_S\frac{1}{r^3}>\;,\\
<~^5D_2|\Delta E|~^5D_2> & = & -\alpha_S\;\frac{4}{15m_cm_b}\;<\frac{1}{r^3}> +
\nonumber \\
&&\biggl(\frac{1}{2m_c^2} - \frac{1}{5m_b^2}\biggr)\;
<-\frac{dV(r)}{rdr}+\frac{8}{3}\;\alpha_S\frac{1}{r^3}>\;,\label{6}\\
<~^3D_2|\Delta E|~^3D_2> & = & -\alpha_S\;\frac{8}{15m_cm_b}\;<\frac{1}{r^3}> +
\nonumber \\
&&\biggl(-\frac{3}{4m_c^2} + \frac{9}{20m_b^2}\biggr)\;
<-\frac{dV(r)}{rdr}+\frac{8}{3}\;\alpha_S\frac{1}{r^3}>\;, \\
<~^5D_2|\Delta E|~^3D_2> & = &
-\alpha_S\;\frac{2\sqrt{6}}{15m_cm_b}\;<\frac{1}{r^3}>
\nonumber \\
&&-\frac{\sqrt{6}}{10m_b^2}\;
<-\frac{dV(r)}{rdr}+\frac{8}{3}\;\alpha_S\frac{1}{r^3}>\;,
\end{eqnarray}

As one can see from eq.(\ref{5}), the S-level splitting is essentially
determined by the $|R_{nS}(0)|$ value, which can be related with the leptonic
decay constants of the S-states ($0^-$, $1^-$). Section 3 is devoted to the
calculation of these constants by different ways.
We note here only, that with enough accuracy, the predictions of different
potential models on the $|R_{1S}(0)|$ value are in an agreement with each other
as well as with predictions in other approaches.

For the 2P, 3P and 3D levels, the mixing matrices of the states with the total
quark spin $S=1$ and $S=0$ have the forms
\begin{eqnarray}
|2P,\;1'^+> & = & 0.294 |S=1> + 0.956 |S=0>\;,\\
|2P,\;1^{+}> & = & 0.956 |S=1> - 0.294 |S=0>\;,
\end{eqnarray}
so that in the $1^+$ state the probability of the total quark spin value
$S=1$ is equal to $w(2P) =0.913$,
\begin{eqnarray}
|3P,\;1'^+> & = & 0.371 |S=1> + 0.929 |S=0>\;,\\
|3P,\;1^{+}> & = & 0.929 |S=1> - 0.371 |S=0>\;,
\end{eqnarray}
so that $w(3P) =0.863$,
\begin{eqnarray}
|3D,\;2'^-> & = & -0.566 |S=1> + 0.825 |S=0>\;,\\
|3D,\;2^{-}> & = & 0.825 |S=1> + 0.566 |S=0>\;,
\end{eqnarray}
so that $w(3D) =0.680$.

\begin{table}[t]
\caption{The Masses (in GeV) of the Bound $(\bar b c$) States below
the Threshold of the Decay into the $BD$ Meson Pair, * is the present paper.}
\label{t8}
\begin{center}
\begin{tabular}{||l|c|c|c||}
\hline
state & *  & \cite{10k} & \cite{6ka}\\
\hline
$1^1S_0$    & 6.253       & 6.264      & 6.314     \\
$1^1S_1$    & 6.317       & 6.337      & 6.355     \\
$2^1S_0$    & 6.867       & 6.856      & 6.889     \\
$2^1S_1$    & 6.902       & 6.899      & 6.917     \\
$2^1P_0$    & 6.683       & 6.700      & 6.728     \\
$2P\; 1^+$  & 6.717       & 6.730      & 6.760     \\
$2P\; 1'^+$ & 6.729       & 6.736      & --        \\
$2^3P_2$    & 6.743       & 6.747      & 6.773     \\
$3^1P_0$    & 7.088       & 7.108      & 7.134     \\
$3P\; 1^+$  & 7.113       & 7.135      & 7.159     \\
$3P\; 1'^+$ & 7.124       & 7.142      & --        \\
$3^3P_2$    & 7.134       & 7.153      & 7.166     \\
$3D\; 2^-$  & 7.001       & 7.009      & --        \\
$3^5D_3$    & 7.007       & 7.005      & --        \\
$3^3D_1$    & 7.008       & 7.012      & --        \\
$3D\; 2'^-$ & 7.016       & 7.012      & --        \\
\hline
\end{tabular}
\end{center}
\end{table}

With the account of the calculated splittings,
the $B_c$ mass spectrum is shown on fig.\ref{f1k} and in table \ref{t8}.

The masses of the $B_c$ mesons have been also calculated in papers of
ref.\cite{k20k}.

As one can see from tables \ref{t2} and \ref{t6}, the place of the 1S-level in
the ($\bar b c$) system ($m(1S) \simeq 6.3$ GeV) is predicted by the potential
models with the rather high accuracy $\Delta m(1S) \simeq 30$ MeV, and
the 1S-level splitting into the vector and pseudoscalar states is about
$m(1^-) - m(0^-) \approx 70$ MeV.
\begin{table}[t]
\caption{The Masses (in GeV) of the Lightest Pseudoscalar $B_c$ and Vector
$B_c^*$ States in Different Models, * is the present paper.}
\label{t6}
\begin{center}
\begin{tabular}{|c|c|c|c|c|c|c|c|}
\hline
state    & * & \cite{8k} & \cite{6ka}& \cite{12k} & \cite{5ka}
& \cite{13k}& \cite{19k,7k} \\
\hline
$0^-$ &6.253&6.249&6.314&6.293&6.270&6.243&6.246\\
$1^-$ &6.317&6.339&6.354&6.346&6.340&6.320&6.319\\
\hline
\hline
state    & \cite{10k} & \cite{14k} & \cite{15k}& \cite{16k} & \cite{17k}
& \cite{18k}& \cite{20k} \\
\hline
$0^-$ &6.264&6.320&6.256&6.276&6.286&--&6.255\\
$1^-$ &6.337&6.370&6.329&6.365&6.328&6.320&6.330\\
\hline
\end{tabular}
\end{center}
\end{table}

\subsection{$B_c$ Meson Masses from QCD Sum Rules.}

Potential model estimates for the masses of the lightest ($\bar b c$)
states are in the agreement with the results of the calculations
for the vector and pseudoscalar ($\bar b c$) states in the framework of the
QCD sum rules \cite{20k,21k,22k}, where the calculation accuracy is lower,
than the accuracy of the potential models, because the results essentially
depend on both the modelling the nonresonant hadronic part of the current
correlator (the continuum threshold) and the parameter of the sum rule
scheme (the number of the moment for the spectral density of the current
correlator or the Borel transformation parameter),
\begin{equation}
m^{SR}(0^-)\approx m^{SR}(1^-) \simeq 6.3\div 6.5\;GeV\;.
\end{equation}

As it has been shown in papers of ref.\cite{23k}, for the lightest vector
quarkonium, the following QCD sum rules take place
\begin{equation}
\frac{f_V^2 M_V^2}{m_V^2- q^2} = \frac{1}{\pi}\;
\int_{s_i}^{s_{th}} \frac{\Im m \Pi_V^{QCDpert}(s)}{s-q^2}\; ds +
\Pi_V^{QCDnonpert}(q^2)\;, \label{4.1}
\end{equation}
where $f_V$ is the leptonic constant of the vector ($\bar b c$) state with
the mass $M_V$,
\begin{eqnarray}
i f_V M_V \epsilon^{\lambda}_\mu e^{ipx} & = &
<0|J_\mu (x)|V(p,\;\lambda)>\;,\label{59s}\\
J_\mu (x) & = & \bar c(x) \gamma_\mu b(x)\;,\label{60s}
\end{eqnarray}
where $\lambda$, $p$ are the $B^*_c$ polarization and momentum, respectively,
and
\begin{eqnarray}
\int d^4x\;e^{iqx} <0|TJ_\mu(x) J_\nu(0)|0> & = & \biggl(-g_{\mu\nu}+
\frac{q_\mu q_\nu}{q^2}\biggr)\;\Pi_V^{QCD} + q_\mu q_\nu\;\Pi_S^{QCD}\;,\\
\Pi_V^{QCD}(q^2) & = & \Pi_V^{QCDpert} + \Pi_V^{QCDnonpert}(q^2)\;,\\
\Pi_V^{QCDnonpert}(q^2) & = & \sum C_i(q^2) O^i\;,
\end{eqnarray}
where $O^i$ are the vacuum expectation values of the composite operators such
as $<m\bar \psi \psi>$, $<\alpha_S\;G_{\mu\nu}^2>$ and so on. The Wilson
coefficients are calculable in the perturbation theory of QCD.
$s_i=(m_c+m_b)^2$ is the kinematical threshold of the perturbative
contribution,  $M_V^2 > s_i$, $s_{th}$ is the threshold of the nonresonant
hadronic contribution, which is considered to be equal to the perturbative
contribution at $s > s_{th}$.

Considering the respective correlators, one can write down the sum rules,
analogous to eq.(\ref{4.1}), for the scalar and pseudoscalar states.

One believes that the sum rule (\ref{4.1}) must rather accurately be valid at
$q^2 <0$.

For the $n$-th derivative of eq.(\ref{4.1}) at $q^2=0$ one gets
\begin{equation}
f_V^2 (M_V^2)^{-n} = \frac{1}{\pi}\;
\int_{s_i}^{s_{th}} \frac{\Im m \Pi_V^{QCDpert}(s)}{s^{n+1}}\; ds +
\frac{(-1)^n}{n!}\;\frac{d^n}{d(q^2)^n} \Pi_V^{QCDnonpert}(q^2)\;, \label{4.4}
\end{equation}
so, considering the ratio of the $n$-th derivative to the $n+1$-th
one, one can obtain the value of the vector $B_c^*$ meson mass. The result of
the calculation depends on the $n$ number in the sum rules (\ref{4.4}),
because of taking into the account both still finite number of the terms
in the perturbation theory expansion and restricted set of the composite
operators.

The analogous procedure can be performed in the sum rule scheme with the
Borel transform, leading to the dependence of the results on the transformation
parameter.

As one can see from eq.(\ref{4.4}), the result, obtained in the framework
of the QCD sum rules, depends on the choices of the values for the hadronic
continuum threshold energy and the current masses of the quarks. Then, this
dependence causes large errors in the estimates of the masses for the lightest
pseudoscalar, vector and scalar  $(\bar b c)$ states.

Thus, the QCD sum rules give the estimates of the quark binding energy in the
quarkonium, and the estimates are in the agreement with the results of the
potential models, but they contain the large parametric uncertainty.

\section{Radiative Transitions in the $B_c$ Family.}

The $B_c$ mesons have no annihilation channels for the decays due to QCD and
electromagnetic interactions. Therefore, the mesons, lying below the threshold
for the production of $B$ and $D$ mesons, will, by cascade way, decay into
the $0^-(1S)$ state by emission of $\gamma$ quanta and $\pi$ mesons.
Theoretical estimates of the transitions between the levels with the emission
of the $\pi$ mesons have uncertainties, and the electromagnetic transitions
are rather accurately calculable.

\subsection{Electromagnetic Transitions.}

The formulae for the radiative E1-transitions have the form \cite{1r,2r}
\begin{eqnarray}
\Gamma(\bar nP_J\to n^1S_1 +\gamma) & = & \frac{4}{9}\;\alpha_{em}\;Q^2_{eff}\;
       \omega^3\;I^2(\bar nP;nS)\;w_J(\bar nP) \;,\nonumber\\
\Gamma(\bar nP_J\to n^1S_0 +\gamma) & = & \frac{4}{9}\;\alpha_{em}\;Q^2_{eff}\;
       \omega^3\;I^2(\bar nP;nS)\;(1-w_J(\bar n P)) \;,\nonumber\\
\Gamma(n^1S_1\to \bar n P_J +\gamma) & = &
\frac{4}{27}\;\alpha_{em}\;Q^2_{eff}\;
       \omega^3\;I^2(nS;\bar n P)\;(2J+1)\;
\nonumber \\&&w_J(\bar n P)\;, \label{7} \\
\Gamma(n^1S_0\to \bar n P_J +\gamma) & = &
\frac{4}{9}\;\alpha_{em}\;Q^2_{eff}\;
       \omega^3\;I^2(nS;\bar n P)\;(2J+1)\;
\nonumber \\&&(1-w_J(\bar n P))\;,\nonumber\\
\Gamma(\bar n P_J\to n D_{J'} +\gamma) & = &
\frac{4}{27}\;\alpha_{em}\;Q^2_{eff}\;
       \omega^3\;I^2(nD;\bar n P)\;(2J'+1)\;
\nonumber \\&&w_J(\bar n P)) w_{J'}(nD)
        S_{JJ'}\;,\nonumber\\
\Gamma(n D_J\to \bar n P_{J'} +\gamma) & = &
\frac{4}{27}\;\alpha_{em}\;Q^2_{eff}\;
       \omega^3\;I^2(nD;\bar n P)\;(2J'+1)\;
\nonumber \\&&w_{J'}(\bar n P)) w_{J}(nD)
        S_{J'J}\;,\nonumber
\end{eqnarray}
where $\omega$ is the photon energy, $\alpha_{em}$ is the electromagnetic
fine structure constant, $w_J(nL)$ is the probability that the spin $S=1$
in the $nL$ state, so that
$w_0(nP)=w_2(nP)=1$, $w_1(nD)=w_3(nD)=1$, and the $w_1(nP)$, $w_2(nD)$
values are presented in the previous section.

The statistical factor $S_{JJ'}$ takes the values \cite{2r}

\begin{center}
\begin{tabular}{ccl}
$J$  & $J'$  & $S_{JJ'}$\\
0 & 1 & 2\\
1 & 1 & 1/2\\
1 & 2 & 9/10\\
2 & 1 & 1/50\\
2 & 2 & 9/50\\
2 & 3 & 18/25
\end{tabular}
\end{center}

The $I(\bar nL;nL')$ value is expressed through the radial wave functions,
\begin{equation}
   I(\bar n L;nL') = |\int R_{\bar n L}(r) R_{nL'}(r) r^3 dr|\;.    \label{7aa}
\end{equation}
For the set of the transitions one obtains
\begin{eqnarray}
 I(1S,2P) & = & 1.568\; GeV^{-1}\;,\nonumber\\
 I(1S,3P) & = & 0.255\; GeV^{-1}\;,\nonumber\\
 I(2S,2P) & = & 2.019\; GeV^{-1}\;,\\
 I(2S,3P) & = & 2.704\; GeV^{-1}\;,\nonumber\\
 I(3D,2P) & = & 2.536\; GeV^{-1}\;,\nonumber\\
 I(3D,3P) & = & 2.416\; GeV^{-1}\;.\nonumber
\end{eqnarray}
\begin{table}[t]
\caption{The Energies (in MeV) and Widths (in keV) of the Electromagnetic
E1-Transitions in the $(\bar b c$) Family.}
\label{tr1}
\begin{center}
\begin{tabular}{||l|c|r|r||}
\hline
transition & $\omega$ & $\Gamma$~~~ & $\Gamma$\cite{10k}~\\
\hline
$2P_2\to 1S_1 +\gamma$ & 426  & 102.9  & 112.6 \\
$2P_0\to 1S_1 +\gamma $& 366  & 65.3  & 79.2 \\
$2P\;1'^+\to 1S_1 +\gamma$ & 412  & 8.1  & 0.1\\
$2P\;1^+\to 1S_1 +\gamma$ & 400  & 77.8  & 99.5\\
$2P\;1'^+\to 1S_0 +\gamma$ & 476  & 131.1  & 56.4 \\
$2P\;1^+\to 1S_0 +\gamma$ & 464  & 11.6 & 0.0 \\
\hline
$3P_2\to 1S_1 +\gamma$ & 817  & 19.2  & 25.8 \\
$3P_0\to 1S_1 +\gamma $& 771  &  16.1 & 21.9 \\
$3P\;1'^+\to 1S_1 +\gamma$ & 807  & 2.5 & 2.1 \\
$3P\;1^+\to 1S_1 +\gamma$ & 796  & 15.3  & 22.1\\
$3P\;1'^+\to 1S_0 +\gamma$ & 871  & 20.1  & -- \\
$3P\;1^+\to 1S_0 +\gamma$ & 860  & 3.1  & -- \\
\hline
$3P_2\to 2S_1 +\gamma$ & 232  & 49.4  & 73.8 \\
$3P_0\to 2S_1 +\gamma $& 186  & 25.5  & 41.2 \\
$3P\;1'^+\to 2S_1 +\gamma$ & 222  & 5.9  & 5.4 \\
$3P\;1^+\to 2S_1 +\gamma$ & 211  & 32.1  & 54.3 \\
$3P\;1'^+\to 2S_0 +\gamma$ & 257  & 58.0 & -- \\
$3P\;1^+\to 2S_0 +\gamma$ & 246  & 8.1  & -- \\
\hline
$2S_1\to 2P_2 +\gamma$ & 159  & 14.8  & 17.7   \\
$2S_1\to 2P_0 +\gamma$ & 219  & 7.7  & 7.8   \\
$2S_1\to 2P\;1'^+ +\gamma$ & 173  & 1.0  & 0.0   \\
$2S_1\to 2P\;1^+ +\gamma$ & 185  & 12.8  & 14.5   \\
\hline
$2S_0\to 2P\;1'^+ +\gamma$ & 138  & 15.9  & 5.2   \\
$2S_0\to 2P\;1^+ +\gamma$ & 150  & 1.9  & 0.0   \\
\hline
\end{tabular}
\end{center}
\end{table}

In eq.(\ref{7}) one uses
\begin{equation}
   Q_{eff}=(m_c Q_{\bar b} - m_b Q_c )/(m_c +m_b)\;,
\end{equation}
where $Q_{c,b}$ are the electric charges of the quarks. For the
$B_c$ meson with the parameters from the Martin potential, one gets
$Q_{eff}=0.41$.

\begin{table}[t]
\caption{The Energies (in MeV) and Widths (in keV) of the Electromagnetic
E1-Transitions in the $(\bar b c$) Family.}
\label{tr2}
\begin{center}
\begin{tabular}{||l|c|r|r||}
\hline
transition & $\omega$ & $\Gamma$~~ & $\Gamma$\cite{10k}\\
\hline
$3P_2\to 3D_1 +\gamma$ & 126  & 0.1 & 0.2 \\
$3P_2\to 3D\;2'^- +\gamma$ & 118  & 0.5  & --\\
$3P_2\to 3D\;2^- +\gamma$ & 133  & 1.5 & 3.2\\
$3P_2\to 3D_3 +\gamma$ & 127 & 10.9  & 17.8 \\
$3P_0\to 3D_1 +\gamma $& 80  & 3.2 & 6.9 \\
$3P\;1'^+\to 3D_1 +\gamma$ & 116  & 0.3 & 0.4 \\
$3P\;1^+\to 3D_1 +\gamma$ & 105  & 1.6  & 0.3 \\
$3P\;1'^+\to 3D\;2'^- +\gamma$ & 108  & 3.5  & -- \\
$3P\;1^+\to 3D\;2^- +\gamma$ & 112  & 3.9 & 9.8\\
$3P\;1'^+\to 3D\;2^- +\gamma$ & 123  & 2.5  & 11.5 \\
$3P\;1^+\to 3D\;2'^- +\gamma$ & 97  & 1.2 & --\\
\hline
$3D_3 \to 2P_2 + \gamma$ & 264  & 76.9  & 98.7  \\
$3D_1 \to 2P_0 + \gamma$ & 325  & 79.7  & 88.6  \\
$3D_1 \to 2P\;1'^+ + \gamma$   & 279  & 3.3  & 0.0  \\
$3D_1 \to 2P\;1^+ + \gamma $  & 291  & 39.2  & 49.3  \\
$3D_1 \to 2P_2 + \gamma $  & 265  & 2.2  & 2.7  \\
$3D\;2'^- \to 2P_2 + \gamma$ & 273  & 6.8  & --   \\
$3D\;2'^- \to 2P_2 + \gamma$ & 258  & 12.2  & 24.7  \\
$3D\;2'^- \to 2P\;1'^+ + \gamma$ & 287  & 46.0  & 92.5  \\
$3D\;2'^- \to 2P\;1^+ + \gamma$ & 301  & 25.0  & --  \\
$3D\;2^- \to 2P\;1'^+ + \gamma$ & 272  & 18.4  & 0.1  \\
$3D\;2^- \to 2P\;1^+ + \gamma$ & 284  & 44.6  & 88.8  \\
\hline
\end{tabular}
\end{center}
\end{table}

For the dipole magnetic transitions one has \cite{2k,1r,2r}
\begin{equation}
\Gamma(\bar n^1S_i\to n^1S_f +\gamma) =  \frac{16}{3}\;\mu^2_{eff}\;\omega^3\;
(2f+1)\;A_{if}^2\;,
\label{10}
\end{equation}
where
$$
A_{if} = \int R_{\bar n S}(r) R_{nS}(r) j_0(\omega r/2) r^2 dr\;,
$$
and
\begin{equation}
\mu_{eff}=\frac{1}{2}\;\frac{\sqrt{\alpha_{em}}}{2m_c m_b}\;
(Q_c m_b - Q_{\bar b} m_c)\;.\label{11}
\end{equation}
Note, in contrast to the $\psi$- and $\Upsilon$-particles, the total width of
the $B_c^*$ meson is equal to the width of its radiative decay into
the $B_c(0^-)$ state.

The electromagnetic widths, calculated under eqs.(\ref{7}),(\ref{10}),
and the frequencies of the emitted photons are presented in tables \ref{tr1},
\ref{tr2}, \ref{tr3}.

Thus, the registration of the cascade electromagnetic transitions in the
$(\bar b c)$ family can be used for the observation of the higher
$(\bar b c)$excitations, having no annihilation channels of the decays.

\subsection{Hadronic Transitions.}

In the framework of QCD the consideration of the hadronic
transitions between the states of the
heavy quarkonium family is built on the basis of the multipole expansion
for the gluon emission by the heavy nonrelativistic quarks \cite{24k}, with
the postcoming hadronization of the gluons, independently of the heavy
quark motion.

\begin{table}[t]
\caption{The Energies (in MeV) and Widths (in keV) of the Electromagnetic
M1-Transitions in the $(\bar b c$) Family.}
\label{tr3}
\begin{center}
\begin{tabular}{||l|r|r|r||}
\hline
transition & $\omega$ & $\Gamma$~~ & $\Gamma$\cite{10k}\\
\hline
$2S_1\to 1S_0 +\gamma$ & 649  & 0.098  & 0.123   \\
$2S_0\to 1S_1 +\gamma$ & 550  & 0.096  & 0.093   \\
$1S_1\to 1S_0 +\gamma$ & 64  & 0.060  & 0.135   \\
$2S_1\to 2S_0 +\gamma$ & 35  & 0.010  & 0.029   \\
\hline
\end{tabular}
\end{center}
\end{table}

In the leading approximation over the velocity of the heavy quark motion,
the action, corresponding to the heavy quark coupling to the external
gluon field,
\begin{equation}
S_{int} = -g\;\int d^4x\;A_\mu^a(x)\cdot j^\mu_a(x)\;,
\end{equation}
can be expressed in the form
\begin{equation}
S_{int}  =  g\;\int dt \;
r^k E_k^a(t,\vec{x})\;\frac{\lambda_a^{ij}}{2} \Psi_n(\vec{r})
\Psi_f^{ji}(\vec{r}) \;K(s_n,f) \;d^3\vec{r}\;,
\end{equation}
where $\Psi_n(\vec{r})$ is the wave function of the quarkonium, emitting the
gluon, $\Psi_f^{ij}(\vec{r})$ is the wave function of the colour-octet
state of the quarkonium, $K(s_n,f)$  respects to the spin factor (in the
leading approximation, the heavy quark spin is decoupled from the interaction
with the gluons).

Then the matrix element for the E1-E1 transition of the quarkonium
$nL_J \to n'L'_{J'} + gg$ can be written in the form
\begin{eqnarray}
M(nL_J \to n'L'_{J'}+gg) & = & 4\pi \alpha_S\;E_k^a E_m^b\;
\cdot \nonumber \\
{}~ & ~ & \int d^3r d^3r'\;r_k r'_m\;G^{ab}_{s_{n'},s_n}(r,r')\;
\Psi_{nL_J}(r) \Psi_{n'L'_{J'}}(r'), \label{h1}
\end{eqnarray}
where $G^{ab}_{s_{\bar n},s_n}(r,r')$ respects to the propagator of the
colour-octet state of the heavy quarkonium
\begin{equation}
G = \frac{1}{\epsilon - H_{Q\bar Q}^c}\;,
\end{equation}
where $H_{Q\bar Q}^c$ is the hamiltonian of the coloured state.

One can see from eq.(\ref{h1}), that the determination of the transition
matrix element depends on both the wave function of the quarkonium and the
hamiltonian $H_{Q\bar Q}^c$.
Thus, the theoretical consideration of the hadronic transitions in the
quarkonium family is model dependent.

In the set of papers of ref.\cite{25k}, for the calculation of the values
such as (\ref{h1}), the potential approach has been developed.

In papers of ref.\cite{26k} it has been shown that nonperturbative conversion
of the gluons into the $\pi$ meson pair allows one to make the consideration
in the framework of the low-energy theorems in QCD, so that this consideration
agrees with the papers, performed in the framework of PCAC and soft pion
technique \cite{27k}.

However, as it follows from eq.(\ref{h1}) and the Wigner-Eckart theorem,
the differential width for the E1-E1 transition allows the representation
in the form \cite{25k}
\begin{equation}
\frac{d\Gamma}{dm^2}(nL_J \to n'L'_{J'} + h) = (2J'+1)\;
\sum_{k=0}^{2} \left\{\begin{array}{ccc}
k& L& L'\\ s & J'& J \end{array}\right\}^2 A_k(L,L')\;, \label{h2}
\end{equation}
where $m^2$ is the invariant mass of the light hadron system $h$, $\{~\}$ are
6j-symbols, $A_k(L,L')$ is the contribution by the irreducible tensor of the
rang, equal to $k=0,\; 1,\; 2$,
$s$ is the total quark spin inside the quarkonium.

In the limit of soft pions, one has $A_1(L,L')=0$.

{}From eqs.(\ref{h1}), (\ref{h2}) it follows, that, with the accuracy by the
difference in the phase spaces, the widths of the hadronic transitions in the
$(Q\bar Q)$ and $(Q\bar Q')$ quarkonia are related by the following expression
\cite{24k,25k}
\begin{equation}
\frac{\Gamma  (Q \bar Q')}{\Gamma (Q \bar Q)} = \frac{<r^2(Q\bar Q')>^2}
{<r^2(Q\bar Q)>^2}\;. \label{h3}
\end{equation}
Then the experimental data on the transitions of $\psi ' \to J/\psi + \pi\pi$,
$\Upsilon ' \to \Upsilon + \pi\pi$, $\psi(3770) \to J/\psi + \pi\pi$
\cite{28k} allow one to extract the values of $A_k(L,L')$ for the transitions
$2S \to 1S + \pi\pi$ and $3D \to 1S + \pi\pi$ \cite{10k}.

The invariant mass spectrum of the $\pi$ meson pair has the universal form
\cite{26k,27k}
\begin{equation}
\frac{1}{\Gamma}\;\frac{d\Gamma}{dm} = B\;\frac{|\vec{k_{\pi\pi}}|}
{M^2}\;(2x^2 -1)^2\;\sqrt{x^2-1}\;, \label{h4}
\end{equation}
where $x=m/2m_\pi$, $|\vec{k_{\pi\pi}}|$ is the $\pi\pi$ pair momentum.

The estimates for the widths of the hadronic transitions in the $(\bar b c)$
family have been made in ref.\cite{10k}. The hadronic transition widths,
having the values comparable with the electromagnetic transition width values,
are presented in table \ref{th1}.
\begin{table}[t]
\caption{The Widths (in keV) of the Radiative Hadronic Transitions in the
$(\bar b c)$ Family.}
\label{th1}
\begin{center}
\begin{tabular}{||c|c||}
\hline
transition & $\Gamma$ \cite{10k} \\
\hline
$2S_0 \to 1S_0 + \pi\pi$ & 50 \\
$2S_1 \to 1S_1 + \pi\pi$ & 50 \\
$3D_1 \to 1S_1 + \pi\pi$ & 31 \\
$3D_2 \to 1S_1 + \pi\pi$ & 32 \\
$3D_3 \to 1S_1 + \pi\pi$ & 31 \\
$3D_2 \to 1S_0 + \pi\pi$ & 32 \\
\hline
\end{tabular}
\end{center}
\end{table}
The transitions in the $(\bar b c)$ family with the emission of $\eta$ mesons
are suppressed by the low value of the phase space.

Thus, the registration of the hadronic transitions in the $(\bar b c)$ family
with the emission of the $\pi$ meson pairs can be used to observe the higher
2S- and 3D-excitations of the basic state.

\section{Leptonic Constant of $B_c$ Meson.}

As we have seen in Section 1, the value of the leptonic constant of the
$B_c$ meson determines the splitting of the basic 1S-state of the
($\bar b c$) system. Moreover, the higher excitations in the ($\bar b c$)
system transform, by cascade way, into the lightest $0^-$ state of $B_c$,
whose widths of the decays are essentially determined by the value of
$f_{B_C}$, too. In the quark models \cite{1a,2a,3a}, used to calculate the
weak decay widths of mesons, the leptonic constant, as the parameter,
determines the broadness of the quark wave packet inside the meson (generally,
the wave function is chosen in the oscillator form), therefore, the practical
problem for the extraction of the value for the weak charged current mixing
matrix element $|V_{bc}|$ from the data on the weak $B_c$ decays can
be only solved at the known value of $f_{B_C}$.

Thus, the leptonic constant $f_{B_C}$ is the most important quantity,
characterizing the bound state of the ($\bar b c$) system.

In the present Section we calculate the value of $f_{B_C}$ in different ways.

To describe the bound states of the quarks requires the use of the
nonperturbative approaches. The bound states of the heavy quarks allow one to
consider simplifications, connected to both large values of the quark masses
$\Lambda_{QCD}/m_Q \ll 1$ and the nonrelativistic quark motion $v \to 0$.
Therefore the value of $f_{B_C}$ can be rather reliably determined in the
framework of the potential models and the QCD sum rules \cite{23k}.

\subsection{$f_{B_C}$ from Potential Models.}

In the framework of the nonrelativistic potential models, the leptonic
constants of the pseudoscalar and vector mesons (see eqs.(\ref{59s}),
(\ref{60s}))
\begin{eqnarray}
<0|\bar c(x) \gamma_\mu b(x)|B_c^*(p,\epsilon)> & = & i f_V\; M_V \;
\epsilon_\mu\; e^{ipx}\;, \\
<0|\bar c(x) \gamma_5 \gamma_\mu b(x)|B_c(p)> & = & i f_P\; p_\mu \;
 e^{ipx}\;,
\end{eqnarray}
are determined by expression (\ref{27s})
\begin{equation}
f_V = f_P = \sqrt{\frac{3}{\pi M_{B_C(1S)}}} R_{1S}(0)\;,\label{3a}
\end{equation}
where $R_{1S}(0)$ is the radial wave function of the 1S-state of the
($\bar b c$) system, at the origin. The wave function is calculated by
solving the Schr\" odinger equation with the different potentials
\cite{2k,5ka,3k,4k,5k,6ka}, in the quasipotential approach \cite{5a} or
by solving the Bethe-Salpeter equation with instant potential and in expansion
up to second order over the quark motion velocity $v/c$ \cite{6a,7a}.

The values of the leptonic $B_c$ meson constant, calculated in the different
potential models and effective Coulomb potential with the "running"
$\alpha_S$ constant, determined in Section 1, are presented in table \ref{t1a}.

\begin{table}[b]
\caption{The Leptonic $B_c$ Meson Constant, Calculated in the Different
Potential Models (the accuracy $\sim 15 \%$), in MeV.}
\label{t1a}
\begin{center}
\begin{tabular}{||c|c|c|c|c|c|c|c||}
\hline
model & Martin & Coulomb & \cite{5ka} & \cite{10k}& \cite{5a} &
\cite{6a,7a}&\cite{8a} \\
\hline
$f_{B_C}$ & 510 & 460 & 570 & 495 & 410 & 600 & 500 \\
\hline
\end{tabular}
\end{center}
\end{table}

Thus, in the approach accuracy, the potential quark models give the $f_{B_C}$
values, being in the good agreement with each other, so that
\begin{equation}
f^{pot}_{B_C} = 500 \pm 80\;\;MeV\;.\label{4a}
\end{equation}

\subsection{$f_{B_C}$ from QCD Sum Rules.}

In the framework of the QCD sum rules \cite{23k}, expressions
(\ref{4.1})-(\ref{4.4}) have been derived for the vector states. The
expressions are considered at $q^2 <0$ in the schemes of the spectral density
moments (\ref{4.4}) or with the application of the Borel transform \cite{23k}.
As one can see from eqs.(\ref{4.1}) - (\ref{4.4}), the result of the QCD sum
rule calculations is determined by not only the physical parameters such as
the quark and meson masses, but also by the unphysical parameters of the
sum rule scheme such as the number of the spectral density moment or the Borel
transformation parameter. In the QCD sum rules, this unphysical dependence
of the $f_{B_C}$ value is caused by that the consideration is yet performed
with the finite number of terms in the expansion of the QCD perturbation
theory for the Wilson coefficients of the unit and composite operators.
In the calculations, the set of the composite operators is also restricted.

Thus, the ambiguity in the choice of the hadronic continuum threshold and the
parameter of the sum rule scheme essentially reduces the reliability of the
QCD sum rule predictions for the leptonic constants of the vector and
pseudoscalar $B_c$ states.

Moreover, the nonrelativistic quark motion inside the heavy quarkonium
$v \to 0$ leads to that the $\alpha_S/v$-corrections to the perturbative
part of the quark current correlators become the most important, where
$\alpha_S$ is the effective Coulomb coupling constant in the heavy quarkonium.
As it has been noted in refs.\cite{7k,23k,9a}, the Coulomb
$\alpha_S/v$-corrections can be summed and represented in the form of the
factor, corresponding to the Coulomb wave function of the heavy quarks, so
that
\begin{equation}
F(v) = \frac{4\pi \alpha_S}{3v}\;\frac{1}{1-\exp(-4\pi \alpha_S/3v)}\;,
\label{5a}
\end{equation}
where $2v$ is the relative velocity of the heavy quarks inside the quarkonium.
The expansion of the factor (\ref{5a}) in the first order over $\alpha_S/v$
\begin{equation}
F(v) \simeq 1-\frac{2\pi \alpha_S}{3v}\;,
\label{6a}
\end{equation}
gives the expression, obtained in the first order of the QCD perturbation
theory \cite{23k}.

Note, the $\alpha_S$ parameter in eq.(\ref{5a}) has to be at the scale of the
characteristic quark virtualities in the quarkonium (see Section 1),
but at the scale of the quark or quarkonium masses, as sometime one does it and
decreases the value of the factor (\ref{5a}).

The choice of the $\alpha_S$ parameter essentially determines the spread of
the sum rule predictions for the $f_{B_C}$ value (see table \ref{t2a})
\begin{equation}
f^{SR}_{B_C} = 160 \div 570\;\;MeV\;.\label{7a}
\end{equation}
\begin{table}[b]
\caption{The Leptonic $B_c$ Constant, Calculated in the QCD Sum Rules,
* is the Scaling Relation, in MeV.}
\label{t2a}
\begin{center}
\begin{tabular}{||c|c|c|c|c|c|c|c|c||}
\hline
model & \cite{9a}& \cite{20k} & \cite{21k} & \cite{22k}& \cite{10a} &
\cite{11a}&\cite{12a}& * \\
\hline
$f_{B_C}$ & 375 & 400 & 360 & 300 & 160 & 300 & 450& 460\\
\hline
\end{tabular}
\end{center}
\end{table}

As one can see from eq.(\ref{7a}), the ambiguity in the choices of the QCD
sum rule parameters leads to the essential deviations of the results from
the $f_{B_C}$ estimates (\ref{4a}) in the potential models.

However, as it has been noted in Section 1,
\begin{description}
\item[i~~)]
the large value of the heavy quark masses $\Lambda_{QCD}/m_Q \ll 1$,
\item[ii~)]
the nonrelativistic heavy quark motion inside the heavy quarkonium $v \to 0$,
and
\item[iii)]
the universal scaling properties of the potential in the heavy quarkonium, when
the kinetic energy of the quarks and the quarkonium state density do not depend
on the heavy quark flavours (see eqs.(\ref{v}) - (\ref{e2})),
\end{description}
allow one to state the scaling relation (\ref{law}) for the leptonic constants
of the S-wave quarkonia
$$
\frac{f^2}{M}\;\biggl(\frac{M}{4\mu}\biggr)^2 = const.
$$

Indeed,
\begin{description}
\item[i~~)]
at $\Lambda_{QCD}/m_Q \ll 1$ one can neglect the quark-gluon condensate
contribution, having the order of magnitude $O(1/m_b m_c)$ (the contribution
into the $\psi$ and $\Upsilon$ leptonic constants is less than 15\%),
\item[ii~)]
at $v\to 0$ one has to take into the account the Coulomb-like
$\alpha_S/v$-corrections in the form of the factor (\ref{5a}), so that the
imaginary part of the correlators for the vector and axial quark currents
has the form
\begin{equation}
\Im m \Pi_V(q^2) \simeq \Im m \Pi_P(q^2) = \frac{\alpha_S}{2}\;q^2\;
\biggl(\frac{M}{4\mu}\biggr)^2\;, \label{8a}
\end{equation}
where
$$
v^2 = 1- \frac{4m_b m_c}{q^2-(m_b-m_c)^2}\;,\;\; v\to 0\;.
$$
\end{description}
Moreover, condition (\ref{e2}) can be used in the specific QCD sum rule scheme,
so that this scheme excludes the dependence of the results on the parameters
such as the number of the spectral density moment or the Borel parameter.

Indeed, for example, the resonance contribution into the hadronic part
of the vector current correlator, having the form
\begin{equation}
\Pi_V^{(res)}(q^2)  =  \int \frac{ds}{s-q^2}\;\sum_n f^2_{Vn} M^2_{Vn}
\delta(s-M_{Vn}^2)\;,
\end{equation}
can be rewritten as
\begin{equation}
\Pi_V^{(res)}(q^2) = \int \frac{ds}{s-q^2}\; s f^2_{Vn(s)}\;\frac{dn(s)}{ds}\;
\frac{d}{dn} \sum_k \theta(n-k)\;.
\end{equation}
where $n(s)$ is the number of the vector S-state versus the mass, so that
\begin{equation}
n(m_k^2) = k\;.
\end{equation}
Taking the average value for the derivative of the step-like function, one gets
\begin{equation}
\Pi_V^{(res)}(q^2) = <\frac{d}{dn} \sum_k \theta(n-k)>\; \int \frac{ds}{s-q^2}
s f^2_{Vn(s)} \frac{dn(s)}{ds}\;,
\end{equation}
and, supposing
\begin{equation}
<\frac{d}{dn} \sum_k \theta(n-k)> \simeq 1\;,
\end{equation}
one can, in average, write down
\begin{equation}
\Im m <\Pi^{(hadr)}(q^2)> = \Im m \Pi^{(QCD)}(q^2)\;,
\end{equation}
so, taking into the account the Coulomb factor and neglecting power
corrections over $1/m_Q$, at the physical points $s_n =M_n^2$ one obtains
\begin{equation}
\frac{f_n^2}{M_n}\; \biggl(\frac{M}{4\mu}\biggr)^2 =
\frac{\alpha_S}{\pi} \; \frac{dM_n}{dn}\;, \label{14a}
\end{equation}
where one has supposed that
\begin{equation}
m_b +m_c \approx M_{B_C}\;,\label{15a}
\end{equation}
and
\begin{equation}
f_{Vn} \simeq f_{Pn} = f_n\;.
\end{equation}
Further, as it has been shown in Section 1, in the heavy quarkonium the value
of $dn/dM_n$ does not depend on the quark masses (see eq.(\ref{e2})),
and, with the accuracy up to the logarithmic corrections, $\alpha_S$ is the
constant value (the last fact is also manifested in the flavour independence
of the Coulomb part of the potential in Cornel model). Therefore, one can
draw the conclusion that, in the leading approximation, the right hand side
of eq.(\ref{14a}) is the constant value, and there is the scaling relation
(\ref{law}) \cite{7k}. This relation is valid in the resonant region,
where one can neglect the contribution by the hadronic continuum.

Note, scaling relation (\ref{law}) is in the good agreement with  the
experimental data on the leptonic decay constants of the $\psi$- and
$\Upsilon$-particles (see table \ref{t7}), for which one has $4\mu/M = 1$
\cite{7k}.

The value of the constant in the right hand side of eq.(\ref{law}) is in the
agreement with the estimate, when we suppose
\begin{equation}
<\frac{dM_{\Upsilon}}{dn}> \simeq
\frac{1}{2}((M_{\Upsilon'}-M_{\Upsilon}) + (M_{\Upsilon''}- M_{\Upsilon'}))\;,
\end{equation}
and $\alpha_S= 0.36$, as it is in Cornel model.

Further, in the limit case of $B$- and $D$-mesons,
when the heavy quark mass is much greater than the light quark mass
$m_Q \gg m_q$, one has
$$
\mu \simeq m_q
$$
and
\begin{equation}
f^2\;M = \frac{16 \alpha_S}{\pi}\;\frac{dM}{dn}\;\mu^2\;.\label{18a}
\end{equation}
Then it is evident that at one and the same $\mu$ one gets
\begin{equation}
f^2\;M = const. \label{19a}
\end{equation}
Scaling law (\ref{19a}) is very well known in EHQT
\cite{13a} for mesons with a single heavy quark ($Q\bar q$),
and it follows, for example, from the identity of the $B$- and $D$-meson
wave functions in the limit, when infinitely heavy quark can be considered
as a static source of gluon field.

In our derivation of eqs.(\ref{18a}) and (\ref{19a}) we have neglected
power corrections over the inverse heavy quark mass.
Moreover, we have used the presentation about the light constituent quark
with the mass, equal to
\begin{equation}
m_q \simeq 330\;\;MeV\;, \label{20a}
\end{equation}
so that this quark has to be considered as nonrelativistic one $v \to 0$,
and the following conditions take  place
\begin{equation}
m_Q +m_q \approx M^{(*)}_{(Q\bar q)}\;,\;\;m_q \ll m_Q\;, \label{21a}
\end{equation}
and
\begin{equation}
f_{V} \simeq f_{P} = f\;.
\end{equation}

In agreement with eqs.(\ref{18a}) and (\ref{20a}), one finds the
estimates\footnote
{In ref.\cite{7k} the dependence of the S-wave state density
$dn/dM_n$ on the reduced mass of the system with the Martin potential
has been found by the Bohr-Sommerfeld quantization, so that at the step from
($\bar b b$) to ($\bar b q$), the density changes less than about 15\%.}
\begin{eqnarray}
f_{B^{(*)}} = 120 \pm 20\;\;MeV\;, \\
f_{D^{(*)}} = 220 \pm 30\;\;MeV\;,
\end{eqnarray}
that is in an agreement with the estimates in the other schemes of the QCD sum
rules \cite{23k,14a}.

Thus, in the limits of $4\mu/M=1$ and $\mu/M \ll 1$, scaling relation
(\ref{law}) is consistent.

The $f_{B_C}$ estimate from eq.(\ref{law}) contains the uncertainty,
connected to the choice of the ratio for the $b$- and $c$-quark masses,
so that (see table \ref{t2a})
\begin{equation}
f_{B_C} = 460 \pm 60\;\;MeV\;. \label{24a}
\end{equation}
In ref.\cite{9a} the sum rule scheme with the double Borel transform has been
used. So, it allows one to study effects, related to the power corrections from
the gluon condensate, corrections due to nonzero quark velocity and nonzero
binding energy of the quarks in the quarkonium.

Indeed, for the set of narrow pseudoscalar states, one has the sum rules
\begin{equation}
\sum_{k=1}^\infty \frac{M_k^4 f_{Pk}^2}{(m_b+m_c)^2 (M_k^2-q^2)} =
\frac{1}{\pi} \int \frac{ds}{s-q^2}\;\Im m \Pi_P(s) + C_G(q^2)\;
<\frac{\alpha_S}{\pi} G^2>\;,\label{a1}
\end{equation}
where
\begin{equation}
C_G(q^2) = \frac{1}{192 m_b m_c}\;\frac{q^2}{\bar q^2}\;
\biggl(\frac{3(3v^2+1)(1-v^2)^2}
{2v^5} \ln \frac{1+v}{1-v} - \frac{9v^4+4v^2+3}{v^4}\biggr)\;,
\end{equation}
and
\begin{equation}
\bar q^2 = q^2-(m_b-m_c)^2\;,\;\;\;v^2=1-\frac{4m_bm_c}{\bar q^2}\;.
\end{equation}
Acting by the Borel operator $L_\tau(-q^2)$ on eq.(\ref{a1}), one gets
\begin{equation}
\sum_{k=1}^\infty \frac{M_k^4 f_{Pk}^2}{(m_b+m_c)^2}\;e^{-M_k^2\tau} =
\frac{1}{\pi} \int {ds}\;\Im m \Pi_P(s) \;e^{-s\tau}+ C'_G(\tau)\;
<\frac{\alpha_S}{\pi} G^2>\;,\label{a2}
\end{equation}
where
\begin{eqnarray}
L_\tau (x) & = & \lim_{n,x\to \infty} \frac{x^{n+1}}{n!}\;
\biggl(-\frac{d}{dx}\biggr)^n\;,\;\;\;n/x=\tau\;,\\
C'_G(\tau) & = & L_\tau(-q^2)\;C_G(q^2)\;.
\end{eqnarray}
For the exponential set in the left hand side of eq.(\ref{a2}),
one uses the Euler-MacLaurin formula
\begin{equation}
\sum_{k=1}^\infty \frac{M_k^4 f_{Pk}^2}{(m_b+m_c)^2}\;e^{-M_k^2\tau} =
\int_{m_n}^{\infty} {dM_k}\;\frac{dk}{dM_k}\;M_k^4 f_{Pk}^2 \;e^{-M_k^2\tau}+
\sum_{k=0}^{n-1} {M_k^4 f_{Pk}^2}\;e^{-M_k^2\tau}+\cdots\;.\label{a3}
\end{equation}
Making the second Borel transform $L_{M_k^2}(\tau)$ on eq. (\ref{a2})
with the account of eq.(\ref{a3}), one finds the expression for the
leptonic constants of the pseudoscalar ($\bar b c$) states, so that
\begin{equation}
f_{Pk}^2 = \frac{2(m_b+m_c)^2}{M_k^3}\;\frac{dM_k}{dk} \;
\bigg\{\frac{1}{\pi} \Im m \Pi_P(M_k^2) + C''_G(M_k^2)\;
<\frac{\alpha_S}{\pi} G^2>\bigg\}\;,\label{a4}
\end{equation}
where we have used the following property of the Borel operator
\begin{equation}
L_\tau (x)\;x^n e^{-bx} \to \delta_+^{(n)}(\tau-b)\;.
\end{equation}
Explicit form for the spectral density and Wilson coefficients can be found in
ref.\cite{9a}.

Expression (\ref{a4}) is in the agreement with the above performed derivation
of scaling relation (\ref{law}).

The numerical effect from the mentioned corrections considers to be not large
(the power corrections are of the order of 10\%), and the uncertainty,
connected to the choice of the quark masses, dominates in the error of the
$f_{B_C}$ value determination (see eq.(\ref{24a})).

Thus, we have shown that, in the framework of the QCD sum rules, the most
reliable estimate of the $f_{B_C}$ value (\ref{24a}) is coming from
the use of the scaling relation (\ref{law}) for the leptonic decay constants of
the quarkonia, and this relation very well agrees with the results of the
potential models.

\section*{Conclusion.}

In the present paper we have considered the spectroscopic characteristics of
the bound states in the ($\bar b c$) system.

We have shown that below the threshold of the ($\bar b c$) system decay
into the $BD$ meson pair, there are 16 narrow states of the $B_c$ meson
family, whose masses can be reliably calculated in the framework of the
nonrelativistic potential models of the heavy quarkonia. The flavour
independence of the QCD-motivated potentials in the region of average
distances between the quarks in the ($\bar b b$), ($\bar c c$) and
($\bar b c$) systems and their scaling properties allow one to find
the regularity of the spectra for the levels, nonsplitted by the spin-dependent
forces: in the leading approximation the state density of the system
does not depend on the heavy quark flavours, i.e. the distances between
the nL-levels of the heavy quarkonium do not depend on the heavy quark
flavours.

We have described the spin-dependent splittings of the ($\bar b c$) system
levels, so, the splittings, appearing in the second order over the inverse
heavy quark masses, $V_{SD}\simeq O(1/m_bm_c)$, with the account of the
variation of the effective Coulomb coupling constant of the quarks
(the interaction is due to relativistic corrections, coming from the one gluon
exchange).

The approaches, developed to describe emission by the heavy quarks, have been
applied to the description of the radiative transitions in the ($\bar b c$)
family, whose states have no electromagnetic or gluonic channels of the
annihilation. The last fact means that, due to the cascade processes with the
emission of photons and pion pairs, the higher excitations take the
transitions into the lightest pseudoscalar $B_c$ meson, decaying by the
weak way. Therefore, the excited states of the ($\bar b c$) system have the
widths, essentially less (by two orders of magnitude) than the widths in the
charmonium and bottomonium systems.

As for the value of the leptonic decay constant $f_{B_C}$, it can be the
most reliably estimated from the scaling relation for the leptonic constants
of the heavy quarkonia, due to the relation, obtained in the framework
of the QCD sum rules in the specific scheme.
In the other schemes of the QCD sum rules, it is necessary to do an
interpolation of the scheme parameters (the hadronic continuum threshold
and the number of the spectral density moment or the Borel parameter) into
the region of the ($\bar b c$) system, so this procedure leads to the essential
uncertainties. The $f_{B_C}$ estimate from the scaling relation agrees with the
results of the potential models, whose accuracy for the leptonic constants
is notably lower. The value of $f_{B_C}$ essentially determines the decay
widths and the production cross sections of the $B_c$ mesons.

The $B_c$ decays have been studied in refs.\cite{7ka,21k,15a}, where it has
been shown, that the $B_c$ lifetime is approximately equal to
\begin{equation}
\tau(B_c) \simeq 0.5\div 0.7\;\;ps\;,
\end{equation}
and the characteristic decay mode, having the preferable signature for the
experimental search, has the branching fraction, equal to
\begin{equation}
Br(B_c^+ \to \psi X) \simeq 17\%\;.
\end{equation}
The decay of $B_c^+ \to \psi \pi^+$, having the branching fraction
\begin{equation}
Br(B_c^+ \to \psi \pi^+) \simeq 0.2\%\;,
\end{equation}
is chosen by CDF Collaboration for the $B_c$ search, when about 20 events are
expected \cite{16a}.

The $B_c$ production at the FNAL and LEP colliders has been studied in
refs.\cite{1b,2b,3b,7ka,16a,17a}, where it has been shown, that the level
of the $B_c$ yield has the order of
$$
\frac{\sigma(B_cX)}{\sigma(\bar b b)} \simeq 2\cdot 10^{-3}\;.
$$

Thus, in the present paper we have described the spectroscopic characteristics
of the $B_c$ mesons, whose search at FNAL and LEP is carrying out, and, as
expected, it will be successfully realized in the nearest future.

\end{document}